\documentclass[doublecol]{epl2}
\usepackage{amssymb,amsmath,bm,graphicx,color}

\title{Spontaneous magnon decays in planar ferromagnet}
\shorttitle{Magnon decays in planar ferromagnet}

\author{V. A. Stephanovich\inst{1}
\and M. E. Zhitomirsky\inst{2,3}  
}

\shortauthor{V. A. Stephanovich and M. E. Zhitomirsky}

\institute{
\inst{1}Institute of Physics, Opole University, Opole, 45-052, Poland\\
\inst{2}Service de Physique Statistique, Magn\'etisme et Supraconductivit\'e,
UMR-E9001 CEA-INAC/UJF, \\ 17 rue des Martyrs, F-38054 Grenoble cedex 9, France \\
\inst{3}Max-Planck-Institut f\"ur Physik Komplexer Systeme, N\"othnitzer str.\
38,  D-01187 Dresden, Germany
}

\pacs{75.30.Ds}{Spin waves}
\pacs{75.10.Jm} {Quantized spin models, including quantum spin frustration}
\pacs{75.50.Dd}{Nonmetallic ferromagnetic materials}

\abstract{We predict that spin-waves in an easy-plane ferromagnet
have a finite lifetime at zero temperature due to spontaneous decays.
In zero field the damping is determined by three-magnon
decay processes, whereas decays in the two-particle channel
dominate in a transverse magnetic field.
Explicit calculations of the magnon damping are performed
in the framework of the spin-wave theory for the
$XXZ$ square-lattice ferromagnet with an anisotropy parameter
$\lambda<1$. In zero magnetic field the decays occur for $\lambda^*<\lambda<1$
with $\lambda^*\approx 1/7$.
We also discuss possibility of experimental observation of the predicted
effect in a number of ferromagnetic insulators.
}

\begin{document}

\maketitle

\section{Introduction}

Zero-point fluctuations are generally present in quantum antiferromagnets, whereas
their role in ferromagnetic  structures is usually considered as minor
\cite{Mattis,Majlis}. Without challenging the fundamental reason behind this conclusion,
we present here an example of a quantum effect specific to anisotropic ferromagnets.
Namely, we predict that magnons in an easy-plane ferromagnet have a finite lifetime
at zero temperature due to spontaneous three-particle decays. In contrast, excitations
in a simple two-sublattice antiferromagnet remain stable at $T=0$ \cite{Harris71,Manousakis91}
and acquire nonzero decay rate only above a threshold magnetic field
\cite{Zhitomirsky99,Kreisel08,Olav08,Syromyatnikov09,Luscher09,Masuda10,Mourigal10}.

The low-frequency dynamics of an isotropic ferromagnet is governed
by conservation of the total spin  ${\bf S}_{\rm tot}$. The ferromagnetic
ground state and excited states can be classified according to
$S^z_{\rm tot}$, where the $z$-axis is chosen parallel to the spontaneous
magnetisation.  In particular, the ground state has $S^z_{\rm tot}=NS$
and every magnon carries an intrinsic quantum number $\Delta S^z =-1$.
Spin conservation translates into conservation of
the total number of spin waves by magnon-magnon interaction.
Particle non-conserving processes, including
spontaneous decays, are forbidden by symmetry for the Heisenberg
ferromagnet.

\begin{figure}[t]
\centerline{
\includegraphics[width = 0.9\columnwidth]{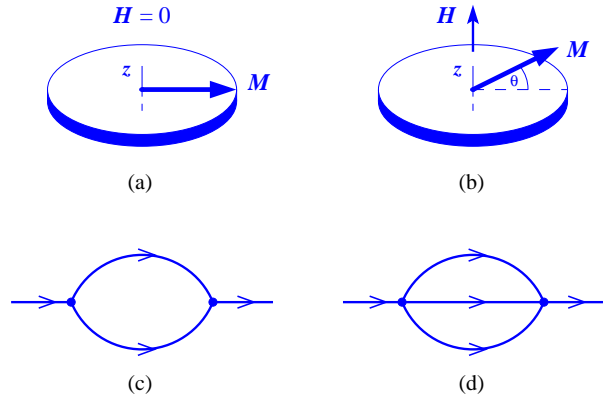}}
\caption{(Colour online) Easy-plane ferromagnet in zero (a)
and in  transverse (b) magnetic field. The self-energy diagrams produced by
two-magnon (c) and three-magnon (d) decay vertices.}
\label{fig:decay_diagrams}
\end{figure}

In the case of an easy-plane ferromagnet the spontaneous magnetisation
is  orthogonal
to the principal $z$-axis, see fig.~\ref{fig:decay_diagrams}a.
The rotation symmetry is completely broken and no value of $S^z$ can be
assigned either to the ground-state or to low-energy excitations:
spin of a spin-wave ceases to exist. Absence of conserved physical quantities
other than the total momentum allows various processes  that change
the magnon number. Among them, spontaneous decays,
fig.~\ref{fig:decay_diagrams}c and \ref{fig:decay_diagrams}d,
determine lifetime of magnetic excitations at zero temperature.
The necessary condition for damping is conservation of energy in an
elementary decay process. In the following we use the spin-wave
theory to study spontaneous magnon decays in an easy-plane ferromagnet
in the two cases: with and without transverse magnetic field.
In the former case both two- (fig.~\ref{fig:decay_diagrams}c)
and three-particle (fig.~\ref{fig:decay_diagrams}d) decay processes are
present, whereas in the latter case only three-particle decays
are compatible with the mirror symmetry $z\to -z$.

The predicted magnon decays at $T\rightarrow 0$ may be observed in easy-plane
ferromagnets such as $\rm K_2CuF_4$
\cite{Harikawa73,Funahashi76,Borovik80}, $\rm CsNiF_3$  \cite{Steiner83} and
$\rm CeRh_3B_2$ \cite{Raymond10}. The second class of physical systems relevant
to our work is the lattice boson models \cite{Scalettar95,Murthy97,Bryant07}.
The equivalence between an $XY$ spin-1/2
ferromagnet and a system of hard-core bosons was established a long time ago
by Matsubara and Matsuda \cite{Matsubara56,Halperin69}.
Bose-Einstein condensates of cold atoms in optical lattices provide
an experimental realization of such bosonic systems \cite{Pethick}.
The analogy between field-induced transitions in quantum magnets and
the Bose-Einstein condensation (BEC) of particles was explored
in many experimental and theoretical studies, for review,
see \cite{Giamarchi08}. The main focus was so far on the critical properties
of magnon BEC.  The dynamical aspects of magnon condensation have
started to attract attention only recently
\cite{Zhitomirsky99,Olav08,Kreisel08,Syromyatnikov09,Luscher09,Mourigal10,Zheludev07,Masuda10}.
An easy-plane ferromagnet
exhibits a quantum critical point in a transverse magnetic field
\cite{Halperin69,Baryakhtar79}, which can be mapped onto the BEC of
magnons \cite{Syromyatnikov07}. Therefore, investigation of
the dynamics of a ferromagnetic  model also helps to clarify
universal dynamical properties of the Bose-Einstein
condensates.

\section{Model}

In this Letter we investigate the zero-temperature dynamics
of the  quantum $XXZ$ ferromagnet
on a square lattice given by the nearest-neighbour Hamiltonian
\begin{equation}
\hat{\mathcal H} = -J\sum _{\langle ij\rangle} \Bigl[ S_i^x S_j^x  +
S^y_i S^y_j + \lambda S^z_i S^z_j\Bigr]  - H\sum_i S^z_i
\label{H0}
\end{equation}
with an easy-plane anisotropy $\lambda<1$ and an arbitrary spin  $S$.
The model (\ref{H0}) describes magnetic properties of
layered ferromagnet $\rm K_2CuF_4$, which has a small
anisotropy $\lambda\approx 0.99$ and an extremely weak
exchange coupling between square planes:
$J'/J<10^{-3}$ \cite{Harikawa73,Funahashi76,Borovik80}.

In zero magnetic field the ferromagnetic moment is oriented arbitrarily in
the $xy$-plane breaking the $SO(2)$ rotation symmetry. In a finite field
the magnetisation tilts away from the easy-plane by an angle $\theta$, see
fig.~\ref{fig:decay_diagrams}b. The mean-field (classical) expression for
the tilting angle $\theta$ at zero-temperature is
\begin{equation}
\sin\theta = \frac{H}{H_c} \ , \quad H_c=4JS(1-\lambda) \ .
\label{theta0}
\end{equation}
For $H>H_c$ local magnetic moments become completely
aligned with the applied field and the $SO(2)$ rotation symmetry
is restored. The transition at $H=H_c$ provides a simple example of the
quantum critical point \cite{Halperin69,Baryakhtar79,Syromyatnikov07}.

We use the spin-wave theory to calculate the damping of
magnetic excitations  in the model (\ref{H0}) at $T=0$.
The derivation for the anisotropic ferromagnet closely follows a similar analysis
for a quantum two-sublattice antiferromagnet in external
field  \cite{Zhitomirsky99,Mourigal10}. The Hamiltonian (\ref{H0})
is rewritten in a tilted coordinate system such that the new
$z'$-axis is directed parallel to the equilibrium magnetisation.
Bosonisation of spin operators is performed in the new frame using the
Holstein-Primakoff transformation \cite{Mattis,Majlis}
\begin{equation}
S^z_i = S - a_i^\dagger a_i \ , \quad S^+_i = (2S-a_i^\dagger a_i)^{1/2}\,a_i \ ,
\label{HP}
\end{equation}
$S_i^\pm = S_i^x\pm iS_i^y$. Square roots are subsequently
expanded to the first order in $1/S$ and
terms up to the fourth order in boson operators $a_i$ and $a_i^\dagger$
are taken into account.

At the harmonic level the boson Hamiltonian  is
diagonalized by the Bogolyubov transformation
\begin{equation}
a_{\bf k}=u_{\bf k}b_{\bf k}+v_{\bf k}b^\dag_{-\bf k} \ .
\end{equation}
The bare magnon energy for $H<H_c$ is
\begin{equation}
\epsilon_{\bf k} =  4JS\sqrt{(1-\gamma_{\bf k})[1-\gamma_{\bf k}(\lambda\cos^2\theta+\sin^2 \theta)]}
\label{Ek}
\end{equation}
with $\gamma_{\bf k} = \frac{1}{2} (\cos k_x + \cos k_y)$, while the Bogolyubov coefficients are
given by
\begin{eqnarray}
&&u_{\bf k} =  \Bigl(\frac{A_{\bf k}+\epsilon_{\bf k}}{2\epsilon_{\bf k}}\Bigr)^{1/2} \ , \quad
v_{\bf k} = -\Bigl(\frac{A_{\bf k}-\epsilon_{\bf k}}{2\epsilon_{\bf k}}\Bigr)^{1/2}\,
\frac{\gamma_{\bf k}}{|\gamma_{\bf k}|} \ , \nonumber \\[0.5mm]
&&A_{\bf k} =  4JS\Bigl[1-\frac{1}{2}\,\gamma_{\bf k} (1 + \lambda \cos^2\theta + \sin^2\theta)\Bigr].
\label{UkVk}
\end{eqnarray}
Zero-point fluctuations in the ground state
$\langle a^\dagger_ia_i\rangle = \sum_{\bf k} v_{\bf k}^2$
are controlled by the parameter $(1-\lambda)\cos^2\theta$.
They are gradually suppressed by the transverse
magnetic field and vanish at $H\geq H_c$.

\begin{figure}[t]
\centerline{
\includegraphics[width = 0.95\columnwidth]{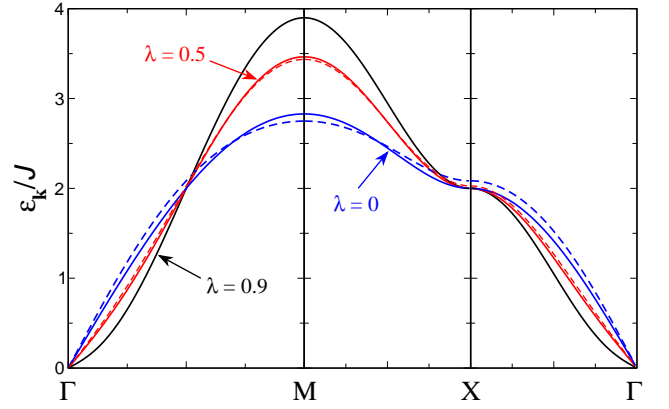}}
\caption{(Colour online) Magnon dispersion of the spin-1/2 $XXZ$ square-lattice ferromagnet
in zero applied field along symmetry directions in the Brillouin zone.
Solid curves show the harmonic energies for three different values of
the anisotropy constant $\lambda$.
Dashed curves correspond to renormalized magnon dispersion. For $\lambda=0.9$
the dashed curve is indistinguishable from the solid line.
}
\label{fig:dispersion}
\end{figure}

The broken $SO(2)$ symmetry in an easy-plane ferromagnet is reminiscent
of the breaking of the $U(1)$ gauge symmetry in the superfluid state
of a Bose gas \cite{Matsubara56,Halperin69}. Accordingly, the energy
spectrum of a quantum $XXZ$ ferromagnet features an acoustic branch
$\epsilon_{\bf k}\sim k$ instead of the `normal' ferromagnetic  dispersion
$\epsilon_{\bf k}\sim k^2$, see fig.~\ref{fig:dispersion}.
Direct expansion of (\ref{Ek}) in small $k$ yields
\begin{eqnarray}
&& \epsilon_{\bf k} \approx ck + \alpha k^3 \ , \quad
c = 2JS\sqrt{1-\lambda}\,\cos\theta \ , \nonumber \\[0.5mm]
&&  \alpha = \frac{c}{8}\,\biggl[
\frac{1}{\cos^2\!\theta\,(1-\lambda)}
- \frac{15+\cos 4\varphi}{12}\biggr]  \ ,
\label{acoustic}
\end{eqnarray}
where $\varphi = \arctan k_y/k_x$.
The excitation spectrum remains gapless up to the critical  field $H_c$.
The magnon velocity is positive in small fields and vanishes at
$H=H_c$. While the critical behaviour is mostly independent of
the second subleading term in the expansion (\ref{acoustic}),
the dynamical properties crucially depend on the convexity of the
acoustic branch \cite{Pitaevskii}.
Considering the anisotropy/field dependence of the coefficient $\alpha(\varphi)$
we can distinguish three different regimes: \\
(i) for $\lambda> 1/4$, $\epsilon_{\bf k}$ is a convex function
of $k$ for $k\to 0$ and for all in-plane directions of the momentum; \\
(ii) for $\lambda^*<\lambda< 1/4$ and $H=0$, the curvature of $\epsilon_{\bf k}$
remains positive near the diagonal $\varphi=\pi/4$
in the Brillouin zone but becomes negative along the principal directions
$\varphi = 0,\pi/2$;\\
(iii) for $\lambda<\lambda^*=1/7$ and $H=0$, $\epsilon_{\bf k}$ has negative curvature
at $k\to 0$, however, there is a threshold magnetic field $H^*$ above which $\alpha(\varphi)$
becomes  positive again:
\begin{equation}
\label{H*}
\frac{H^*}{H_c} = \sqrt{\frac{1-7\lambda}{7(1-\lambda)}} \ .
\end{equation}
A similar change in the convexity of the acoustic branch
in a finite magnetic field takes place in quantum antiferromagnets
\cite{Zhitomirsky99}.

In the isotropic ferromagnet ($\lambda=1$) the expression for the energy
of a single-magnon excitation $\epsilon_{\bf k} = 4JS(1-\gamma_{\bf k})$
is exact, whereas for $\lambda<1$ the harmonic result (\ref{Ek}) is only
approximate. We have calculated the lowest-order Hartree-Fock correction
to (\ref{Ek}) in zero field determined by quartic magnon terms
\cite{Oguchi60}. Results are shown in fig.~\ref{fig:dispersion} by dashed
lines. The quantum correction is momentum dependent
shifting in opposite directions the acoustic branch and the high-energy
magnons. For $\lambda=0$  and $S=1/2$, the  magnon velocity is increased
by 11\%, which is comparable to the 16\% enhancement for the spin-1/2
 square-lattice Heisenberg antiferromagnet \cite{Manousakis91}.
Accordingly, the change in the curvature of the acoustic branch
takes place at $\lambda^*\approx 0.24$ rather than at 1/7.
For $\lambda=0.5$, the quantum renormalization
does not exceed 3--4\% for any $\bf k$.
In the following we disregard completely the real part
of the quantum correction to the spectrum of the $XXZ$ ferromagnet
and focus on the magnon damping given by $\textrm{Im}\,\epsilon_{\bf k}$.

\section{Two-magnon decays}

In the canted magnetic structure at $0<H<H_c$, the principal nonlinear
interaction is provided by the cubic term:
\begin{equation}
\hat{\mathcal H}_3 = J\sqrt{\frac{S}{8}}\,(1-\lambda)\sin 2\theta
\sum_{i,j} a_i^\dag a^{_{}}_i (a^{_{}}_j+a_j^\dag)  \ .
\label{H3}
\end{equation}
After transforming to Bogolyubov bosons one obtains two types of cubic
vertices \cite{Zhitomirsky99}. The magnon damping at $T=0$ is determined by
the two-magnon decay vertex expressed in our case as
\begin{eqnarray}
&& \hat{V}_3\:= \: \frac {1}{2 \sqrt{N}}
\sum_{{\bf k},{\bf q}} V_{\bf k}({\bf q})
\bigl(b^\dag_{\bf q} b^\dag_{{\bf k}-{\bf q}} b^{_{}}_{\bf k} + \textrm{h.\,c.}
\bigr) \ ,
\label{V3} \\
&& V_{\bf k}({\bf q}) = \frac{H\cos\theta}{\sqrt{2S}}\,\bigl[
\gamma_{\bf k}(u_{\bf k}+v_{\bf k})(u_{\bf q}v_{{\bf k}-{\bf q}}
+ v_{\bf q}u_{{\bf k}-{\bf q}})
\nonumber \\ & &
\phantom{V_{{\bf k},{\bf q}} = } \mbox{} + \gamma_{\bf q}
(u_{\bf q}+v_{\bf q}) (u_{\bf k}u_{{\bf k}-{\bf q}} + v_{\bf k}v_{{\bf k}-{\bf q}})
\nonumber \\[0.5mm] & &
 \phantom{V_{{\bf k},{\bf q}} = } \mbox{} + \gamma_{{\bf k}-{\bf q}}(u_{{\bf k}-{\bf q}}+v_{{\bf k}-{\bf q}})
(u_{\bf k}u_{\bf q}+v_{\bf k}v_{\bf q})\bigr] \ .  \nonumber
\end{eqnarray}
The decay vertex vanishes at $H=0$ and $H\geq H_c$.
In the high-field phase this is due to the spin conservation.
In zero field, magnons have no well-defined spin but
still preserve the odd parity under $z\to -z$.
The parity conservation forbids the two-particle decays in zero field,
though the  three-particle decay processes are still possible.

\begin{figure}[t]
\centerline{\includegraphics[width=0.9\columnwidth]{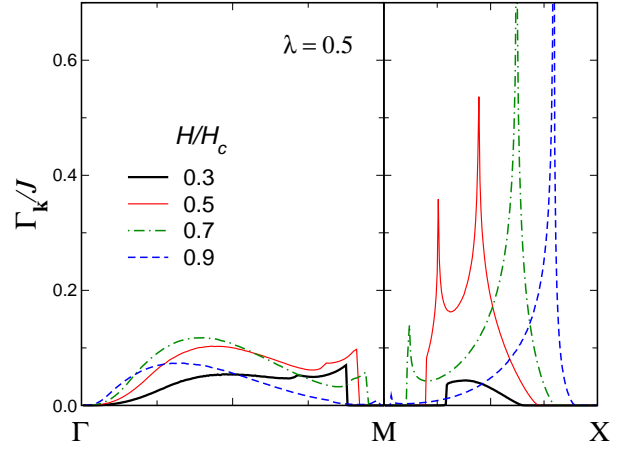}}
\caption{(Colour online) Magnon damping in the two-particle decay channel
for the $XXZ$-ferromagnet with $\lambda=0.5$
in transverse magnetic field.
}
\label{fig:Gk2}
\end{figure}

In the Born approximation, the magnon decay rate in the two-particle channel
is given by the imaginary part of the diagram in fig.~\ref{fig:decay_diagrams}c:
\begin{equation}
\Gamma_{\bf k}=\frac{\pi}{2} \sum_{\bf q}\, V^2_{\bf k}({\bf q})\, 
\delta(\epsilon_{\bf k}-\epsilon_{\bf q}-\epsilon_{\bf k -\bf q}) \ .
\label{Gk2}
\end{equation}
Since $V_{\bf k}({\bf q})=O(JS^{1/2})$ the decay rate (\ref{Gk2}) is independent
of the spin value $\Gamma_{\bf k} = O(J)$.

Spontaneous two-particle decays are allowed if the energy conservation
condition
\begin{equation}
\label{E2}
\epsilon _{\bf k} = \epsilon_{\bf q} + \epsilon_{{\bf k}-{\bf q}}
\end{equation}
is satisfied for a given magnon dispersion. For the acoustic mode
(\ref{acoustic}) solutions of eq.~(\ref{E2}) exist only for positive
values of the coefficient $\alpha$ \cite{Mourigal10,Pitaevskii}.
Following the preceding analysis, we conclude that the
low-energy magnons in the model (\ref{H0}) are kinematically unstable
for $\lambda >1/7$ already in vanishingly small field. Expanding the vertex (\ref{V3})
in small momenta we explicitly obtain for the acoustic mode
\begin{equation}
\Gamma_{\bf k} = \frac{3J}{16\pi} \tan^2\!\theta\; \sqrt{\frac{c}{6\alpha}}
\; k^3 \ .
\label{Gklow}
\end{equation}
Scaling $\Gamma_{\bf k}\propto k^3$ is characteristic
for two-dimensional models \cite{Zhitomirsky99,Kreisel08}.
The decay rate for a 3D anisotropic ferromagnet behaves
similar to  the phonon damping in the superfluid $^4$He:
$\Gamma_{\bf k}\propto k^5$ \cite{Pitaevskii}.

Away from the $k\to 0$ limit, a part of the Brillouin zone with unstable magnons
can be determined numerically by solving (\ref{E2}) for various
incoming momenta. For $\lambda > \lambda^*$ the decay region
occupies a finite area already at $H\to 0$ and spreads out quickly over the entire
Brillouin zone with increasing magnetic field. The magnon decay rate for
arbitrary momenta is obtained by numerical integration of eq.~(\ref{Gk2}).
$\Gamma_{\bf k}$ along two symmetry directions in the Brillouin zone
is shown in fig.~\ref{fig:Gk2} for   $\lambda = 0.5$ and several values of
the transverse magnetic field.
The decay rate is strongest at intermediate fields $H\sim 0.5H_c$, though
it remains significant up to $H=0.9H_c$. The decrease of $V_{\bf k}({\bf q})$ as
$H\to H_c$ is partly compensated by a growing phase-space volume of solutions
of eq.~(\ref{E2}). Jumps and peaks in the momentum dependence of
$\Gamma_{\bf k}$ seeing in fig.~\ref{fig:Gk2} are respectively produced
by the decay thresholds and the logarithmic singularities in the two-magnon
density of states of a 2D model \cite{Chernyshev06}.

A finite lifetime of low-energy spin waves in an easy-plane ferromagnet
in applied field was briefly discussed in \cite{Baryakhtar79}.
Overall, physics of two-particle decays in the
canted ferromagnetic state is quite similar to the field-induced magnon
decays in quantum antiferromagnets
\cite{Zhitomirsky99,Olav08,Kreisel08,Syromyatnikov09,Luscher09,Mourigal10}.
In particular, the asymptotic expression for $\Gamma_{\bf k}$
(\ref{Gklow}) remains valid both in the present case and
for the square-lattice Heisenberg antiferromagnet.
Also, the  self-consistent treatment of the decay processes
beyond the Born approximation removes the 2D Van Hove singularities
in $\Gamma_{\bf k}$ \cite{Mourigal10}.
Away from peaks, the decay rate is not significantly modified by
higher-order processes and the Born approximation results presented in
fig.~\ref{fig:Gk2} may serve as a guidance for future experimental tests.

\begin{figure}[t]
\centerline{\includegraphics[width=0.6\columnwidth]{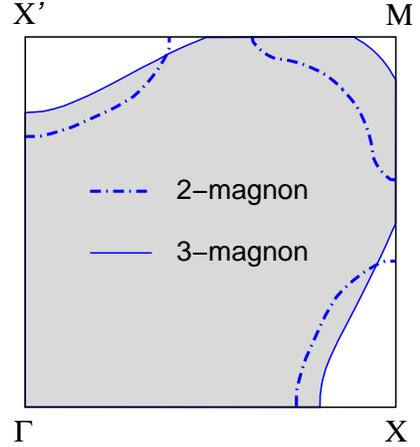}}
\caption{(Colour online)
Three-magnon decay region (shaded area) and kinematic
threshold for two-particle decays (dot-dashed line)
in the first quadrant of the Brillouin zone for the $XXZ$
ferromagnet ($\lambda=0.5$) in zero applied field.
}
\label{fig:decay_region}
\end{figure}

\section{Three-magnon decays}

We now turn to zero-field magnon decays, which
distinguish anisotropic ferromagnets from collinear antiferromagnets.
At $H=0$ the two-magnon decays (\ref{V3}) are forbidden by symmetry.
Still, for $\lambda>\lambda^*$ the energy conservation condition (\ref{E2})
can be satisfied in a certain region in the momentum space.
An example of the kinematic threshold for two-particle decays
is shown in fig.~\ref{fig:decay_region} for $\lambda=0.5$ and $H=0$.
As was discussed by Harris and co-workers \cite{Harris71},
if magnons are stable against two-particle decays
in the entire Brillouin zone, then three-particle decay processes
\begin{equation}
\label{E3}
\epsilon _{\bf k} = \epsilon_{\bf p} + \epsilon_{\bf q}
+ \epsilon_{{\bf k}-{\bf p}-{\bf q}}
\end{equation}
are also energetically forbidden.  Here, we encounter an opposite situation:
conservation of energy and momentum for the
two-magnon processes (\ref{E2}) implies that decays in the three-particle
channel (\ref{E3}) are possible as well  \cite{Maris77}. To our knowledge,
quantum three-particle  decays have not been systematically studied
in the literature. Therefore, apart from specific model results we
will also discuss  below a few general aspects of three-particle decay processes.

The three-magnon decay region in the case of $\lambda=0.5$ and $H=0$
is shown in fig.~\ref{fig:decay_region}  by shaded area.
Its boundary can be determined in a way similar to the two-particle
decay threshold boundary \cite{Pitaevskii,Chernyshev06}.
In particular, velocities of three created  quasiparticles
are equal to each other at the decay threshold.
In fact, a stronger condition of equal momenta for the decay products
holds for the  model dispersion (\ref{Ek}):
\begin{equation}
\epsilon _{\bf k} = 3\epsilon_{{\bf k}/3} \ .
\end{equation}
The three-particle decays typically occur in a larger part of
the Brillouin zone, which encloses the two-magnon decay boundary.

The interaction part responsible for three-particle decays is given by
\begin{equation}
\hat{V}_4= \frac {1}{6N}
\sum_{{\bf k},{\bf p},{\bf q}} U_{\bf k}({\bf p,q})
\bigl(b^\dag_{\bf p} b^\dag_{\bf q} b^\dag_{{\bf k}-{\bf p}-{\bf q}}
b^{_{}}_{\bf k} + \textrm{h.\,c.} \bigr) \ .
\label{V4}
\end{equation}
Derivation of the vertex function $U_{\bf k}({\bf p,q})$
follows the same route as the preceding analysis of the cubic terms. We skip,
therefore,  further technical details and avoid presenting a
cumbersome expression for $U_{\bf k}({\bf p,q})$.
The magnon damping is given by
the imaginary part of the diagram in fig.~\ref{fig:decay_diagrams}d:
\begin{equation}
\Gamma_{\bf k}=\frac{\pi}{6} \sum_{\bf p,q}\,   U^2_{\bf k}({\bf p,q})\,
\delta(\epsilon_{\bf k}-\epsilon_{\bf p}-\epsilon_{\bf q}
- \epsilon_{{\bf k} -{\bf p}-{\bf q}}) \ .
\label{Gk3}
\end{equation}
Since $U_{\bf k}({\bf p,q})=O(J)$, the decay rate (\ref{Gk3}) scales
as $\Gamma_{\bf k}\sim J/S$ representing a higher-order quantum effect
compared to the two-magnon damping (\ref{Gk2}).

\begin{figure}
\centerline{
\includegraphics[width=0.9\columnwidth]{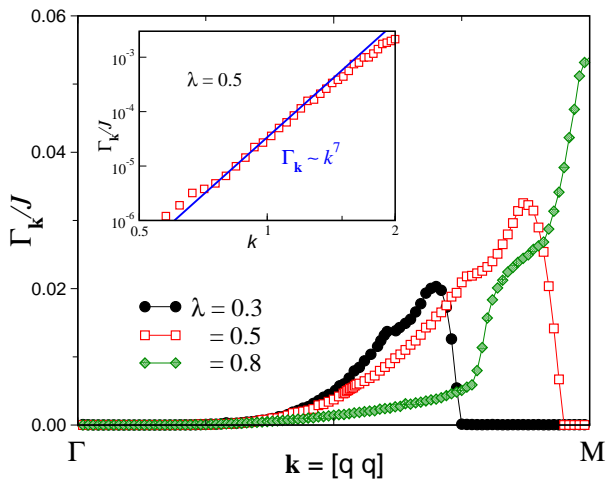}}%
\caption{(Colour online) Zero-field magnon decay rate for the
$XXZ$ spin-1/2 ferromagnet along the $\Gamma$-M line in the Brillouin zone.
The inset: the low-energy asymptote  of $\Gamma_{\bf k}$ on a double-log scale;
the  analytical result (full line)
versus numerical data for $\lambda=0.5$ (squares).
}
\label{fig:Gk3}
\end{figure}

The momentum dependence of $\Gamma_{\bf k}$ along the diagonal
in the Brillouin zone is shown on the main panel of fig.~\ref{fig:Gk3}
for three different values of the anisotropy parameter $\lambda$.
Because of the spin dependence of eq.~(\ref{Gk3}), we used
$S=1/2$ for illustration. For other values of spin $\Gamma_{\bf k}$
is further reduced by a factor $1/(2S)$.
The decay rate behaves linearly  $\Gamma_{\bf k} \sim \Delta k$
in the vicinity of the decay threshold boundary,
whereas the Van Hove singularity
inside the continuum yields a nonanalytic contribution
$\Gamma_{\bf k} \sim \Delta k \ln|\Delta k|$, which shows up
as a shoulder-type feature
on all curves $\Gamma_{\bf k}$ in fig.~\ref{fig:Gk3}.
For both types of anomalies  one finds
 an extra factor $\Delta k$ compared to the behaviour of
the two-particle decay rate \cite{Chernyshev06}.

At small momenta the interaction vertex (\ref{V4})
exhibits a complicated non-analytic behaviour, similar to the momentum
dependence of spin-wave interactions in quantum antiferromagnets
\cite{Harris71,Kopietz90}.
In particular, $U_{\bf k}({\bf p,q})$ can have a finite limiting value
depending on the order
in which ${\bf k}$, ${\bf p}$ and $\bf q$ are taken to 0. Nevertheless,
for the on-shell processes (\ref{E3}) the vertex acquires a normal hydrodynamic
form \cite{Pitaevskii}
\begin{equation}
U_{\bf k}({\bf p,q}) \propto \sqrt{kpqk'} \ , \quad k'=|{\bf k-p-q}|
\label{hydro}
\end{equation}
and vanishes as ${\bf k}\to 0$. Substituting  (\ref{hydro}) into eq.~(\ref{Gk3})
we obtain the power-law
asymptote for the decay rate at low energies: $\Gamma_{\bf k}\propto k^7$.
The derived power-law behaviour is checked on the inset of fig.~\ref{fig:Gk3}.
The deviation between the straight line and  the data points at
small $k$ is due to the representation of the delta-function by a
finite-width Lorentzian. In the three-dimensional case the power-law
exponent changes to $n=11$. Both cases are included in the general
expression
\begin{equation}
\Gamma_{\bf k} \propto k^{4D-1} \ .
\end{equation}
This is again to be compared with a
magnon damping in the  two-particle channel: $\Gamma_{\bf k}\sim k^{2D-1}$
\cite{Kreisel08,Chernyshev06}.

\begin{figure}
\centerline{
\includegraphics[width=0.95\columnwidth]  {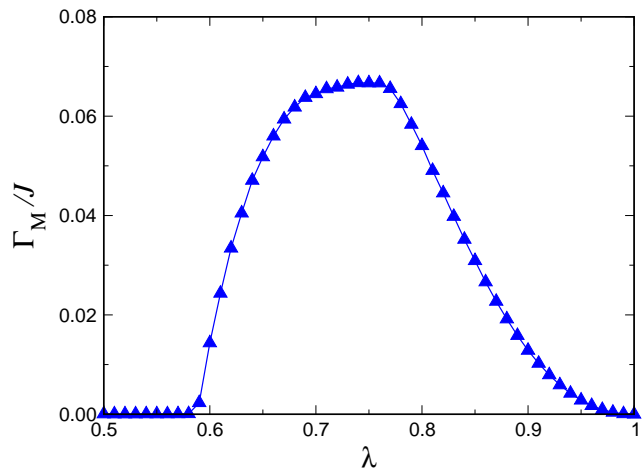}}
\caption{(Colour online) Zero-field magnon decay rate at the M-point
[${\bf k}=(\pi,\pi)$]  in the Brillouin zone
versus the anisotropy parameter $\lambda$.
}
\label{fig:GM}
\end{figure}

An interesting aspect of the obtained results is
the dependence of $\Gamma_{\bf k}$ on anisotropy. At small momenta
the damping decreases with  an increase of  $\lambda$ following the
suppression of the decay vertex $U_{\bf k}({\bf p,q})$.
In the vicinity of the Brillouin zone boundary the tendency is reversed
and $\Gamma_{\bf k}$ is primarily determined by the expanding decay region and
a growing phase-space volume of the decay solutions of eq.~(\ref{E3}).
Eventually,  decays disappear completely at  $\lambda=1$.
The details of such a non-monotonous dependence are shown in fig.~\ref{fig:GM}
for the magnon damping at the M-point.
Summarizing the above results we conclude that the easy-plane anisotropy in the range
$\lambda = 0.5$--0.85 is most favorable for an experimental observation
of the three-particle decays.

In a finite magnetic field (fig.~\ref{fig:decay_diagrams}b) the decay rate is given
by a sum of contributions from different decay channels.
Therefore, for not too small $H$, the magnon lifetime is dominated by
two-particle decays. Still, three-magnon decay processes, being less
restrictive on the value of the initial momentum, will promote decays
beyond the two-particle decay region. Higher-order two-magnon processes,
such as a self-energy insertion in the
inner lines of the diagram in fig.~\ref{fig:decay_diagrams}c,
can be regarded as an effective renormalization of the three-particle
decay vertex $\hat{V}_4$ and have a similar effect on the promotion
of magnon decays \cite{Mourigal10}. As a result, magnons in the entire
Brillouin zone acquire finite lifetimes. Note,
that such a behaviour is not entirely universal.
For the triangular antiferromagnet
two- and three-magnon decay regions coincide and the magnon
damping occurs only in a part of the Brillouin zone \cite{Chernyshev06}.

\section{Conclusions}

We have theoretically investigated zero-temperature magnon decays in an easy-plane
ferromagnet with the exchange anisotropy. The phenomenon of spontaneous magnon
decays depends only on symmetry (absence of spin conservation)
and kinematics (curvature of the spectrum). Hence,  decays should be also present
in ferromagnets with the single-ion anisotropy and different lattice structures.
The most significant damping is produced by two-particle decays, which
are induced by a transverse magnetic field $H<H_c$. In regard to a possible
experimental observation of the predicted effects, square-lattice
ferromagnet $\rm K_2CuF_4$ with $\lambda \approx 0.99$ is perhaps too
isotropic  to have a sizable magnon damping.
Quasi-one-dimensional ferromagnet
$\rm CsNiF_3$ \cite{Steiner83} has a moderate single-ion anisotropy
and provides a good experimental system to test the predicted effects.

Spontaneous three-particle decays is the only damping mechanism in collinear
magnetic structures. For the model case studied here the corresponding decay rate
is somewhat small $\Gamma_{\bf k}/J\leq 0.07$ ($S=1/2$). However,
the significance of three-particle decays extends beyond the present model.
In a collinear antiferromagnet, three-magnon decays are usually impossible due to
the kinematic constraint \cite{Harris71}. However, sizable frustration may significantly
modify the excitation energy $\epsilon_{\bf k}$.
In particular, if an acoustic branch has an upward
curvature at $H=0$, then the energy conservation will allow three-particle decays.
This scenario is realized in the vicinity of the quantum Lifshitz point.
A typical zero-temperature transition in frustrated spin models is
a second-order transition between commensurate and incommensurate antiferromagnetic
structures. At the transition point, the magnon velocity vanishes at least in
one direction in the momentum space
producing a positive curvature of  $\epsilon_{\bf k}$.
Hence, three-particle decays are present
in the commensurate  state, whereas two-particle decays
appear in the incommensurate  phase \cite{Chernyshev06}.

\acknowledgments
We acknowledge stimulating discussions with M. Mourigal and S. Raymond.
We are grateful to M. Gvozdikova for help with the numerical integration
and to A. Chernyshev for careful reading of the manuscript.
Part of this work was performed within the Advanced Study Group Program on
``Unconventional Magnetism in High Fields'' at the Max-Planck Institute for
the Physics of Complex Systems.

\end{document}